\documentclass[aps,prd,onecolumn,superscriptaddress,amsmath,amssymb]{revtex4-2}
\usepackage{orcidlink}
\usepackage{hyperref}

\usepackage{graphicx}
\usepackage{dcolumn}
\usepackage{bm}
\usepackage{hyperref}
\usepackage[mathlines]{lineno}
\usepackage{xcolor}
\usepackage{dsfont}

\usepackage{booktabs}

\usepackage{gensymb}
\usepackage{enumitem}
\usepackage{multirow}
\usepackage{booktabs}
\usepackage{array}
\newcolumntype{P}[1]{>{\centering\arraybackslash}p{#1}}
\usepackage{amssymb}
\usepackage{adjustbox}
\usepackage{orcidlink}

\usepackage[normalem]{ulem}
\usepackage{amsmath}

\newcommand{\iu}{\mathrm{i}\mkern1mu}
\newcommand{\du}{\mathrm{d}}

\newcommand {\cgamma} {\tilde{\gamma}}

\begin{document}

\title{Stability and Formation of Solitonic Boson Stars}

\author{Gareth Arturo Marks \orcidlink{0009-0003-3160-9337}}

\affiliation{DAMTP, Centre for Mathematical Sciences,
University of Cambridge, Wilberforce Road, Cambridge CB3 0WA, UK}

\email{gam54@cam.ac.uk}

\begin{abstract}
We extend a previous study of the radial stability of solitonic boson stars by nonlinearly evolving them under aspherical perturbations, and by studying their formation from the collapse of scalar clouds.
We focus in particular on models previously identified as radially stable, despite having positive binding energy, which ordinarily suggests that dispersion of all scalar matter is energetically preferred. A mode analysis of our 3+1 evolutions produces no evidence for the presence of a non-radial instability in any radially stable model, including those with positive binding energy.   
However, we do find evidence that such models do not generically form via gravitational cooling.
We also find that it is possible to form another kind of compact object, which undergoes long-lasting radial oscillations and hence cannot be described as a stationary boson star.
These objects may be characterized, we suggest, by energetically favored multi-oscillating solutions to the Einstein-Klein-Gordon equations, present in certain regions of the parameter space.
\end{abstract}

\maketitle

\section{Introduction}
The question of whether exotic compact objects (ECOs)--- strongly gravitating bodies beyond the current paradigm of black holes (BHs), neutron stars and white dwarfs--- exist in nature remains an important open problem in astrophysics.
The possibility of astrophysical ECOs is motivated by issues ranging from the unknown nature of dark matter to theoretical issues with classical BHs, such as their curvature singularity and the information paradox \cite{Mathur:2009ip}.
As we prepare to enter a new era of gravitational-wave astronomy with next-generation detectors on the horizon~\cite{abac2025sciencET, Evans:2023euw,amaroseoane2017lisa}, prospects for probing  gravitational signatures associated with ECOs across a wide range of frequencies have never been better. However, many proposed ECO models such as gravastars \cite{Mazur:2001fv} have proven challenging to evolve via numerical relativity (NR), limiting the source-modeling that can be done in these cases.

Boson stars (BSs) present a way to evade these challenges. 
These can be viewed as a self-gravitating configuration of a Bose-Einstein condensate, and arise mathematically from the mere coupling of general relativity to a massive complex scalar field \cite{Kaup:1968, Ruffini:1969qy}. 
As numerically tractable models for ECOs, BSs have been shown to mimic a broad range of BH phenomenology in appropriate regions of the parameter space \cite{Torres:2000dw, Guzman:2005bs, Amaro_Seoane_2010,Rosa_2022sup, Rosa_2022_mim, Rosa_2023}. BSs have also been proposed as candidate objects for dark matter haloes \cite{Sin_1994, Lee:1995af, Schive_2014, Matos:2023usa, Mourelle_24, Mourelle_25}.
Binary mergers consisting of two BSs
\cite{Bezares_2022, Helfer_2022, Evstafyeva:2023kfg, Siemonsen_2023, Ge:2025btw, Ge:2026wzh, Ge:2026fki} and a BS-BH pair \cite{Cardoso_2022, Zhong_2023, Marks:2026xvo, Ning:2026qxs} have received significant attention in NR, allowing for high-precision analysis of their gravitational signatures~\cite{Evstafyeva:2024qvp, Pompili:2025cdc, Evstafyeva:2026juq}.

A class of BSs with particularly rich phenomenology can be obtained by using the so-called \textit{solitonic} potential, which introduces a degenerate vacuum state for the scalar field \cite{Lee:1987}. 
The presence of this additional vacuum can greatly raise the compactnesses that can be achieved in the perturbatively stable regime \cite{Boskovic:2021nfs, Collodel:2022jly}. When the field amplitude at which this vacuum resides is sufficiently small, these families contain \textit{ultracompact objects} (UCOs), solutions so compact that they support a pair of light rings, one stable and one unstable \cite{Cunha:2017qtt,Cunha:2020azh}. 
Such models can thus be regarded as true black hole mimickers, while facilitating evolution in full NR using established frameworks.
For this reason, they have been used in previous numerical studies aimed at determining the efficiency of a conjectured instability related to the presence of the stable light ring \cite{Keir:2014oka, Cunha:2022gde, Marks_2025,Marks:2023ipa, Evstafyeva:2025mvx, Staelens:2025wom}. 
More recently, general-relativistic-magnetohydrodynamic simulations of solitonic BS spacetimes have been used to show that these objects can produce effective shadows \cite{jaramillo2026imitationgamerevolutionsqstar}.

For any ECO model to be astrophysically plausible, it must satisfy two conditions: it must be dynamically stable on suitably long timescales, and it must have a feasible formation mechanism. 
The first nonlinear studies of BS stability were performed in spherical symmetry \cite{Seidel_Suen_1990,Balakrishna_1998}. 
Later, these were supplemented with full 3+1 evolutions \cite{Guzman_2004} and a careful exploration of the dynamical fate of perturbatively unstable solutions \cite{Guzman_2009}. 
Numerical evolutions have also been compared to the results of radial perturbation theory, which in the case of a simple Klein-Gordon potential possibly supplemented with a repulsive quartic self-interaction, predicts a single radially stable branch \cite{Gleiser:1988ih, Kain:2021rmk}.
In all cases considered, linear radial perturbation theory proved sufficient to determine the nonlinear stability of mini BSs both in and out of spherical symmetry: models on the perturbatively stable branch always exhibit long-term dynamical stability, and models on the unstable branch always fail to do so. 
Furthermore, for unstable models, the character of the instability depends on the sign of the \textit{binding energy}--- a quantity measuring the work required to assemble the BS configuration. Unstable models with negative binding energy migrate back to the stable branch or collapse to BHs depending on the type of perturbation added, while those with positive binding energy disperse, the scalar matter escaping to spatial infinity.
The formation of BSs has also been shown to be possible through a process of gravitational cooling \cite{Seidel:1993zk}, and has received attention from the perspective of cosmological perturbation theory in Ref.~\cite{Miyauchi:2025apw}.

In the case of solitonic BSs, we partially addressed the issue of stability in our earlier study \cite{Marks_2025_CP}, which this work extends.
There, we showed that the dynamical fate of solitonic BSs, when nonlinearly evolved in spherical symmetry, could be predicted entirely by the results of linear perturbation theory, even when perturbations were manually prescribed.
Notably, this included UCOs as well as BSs with positive binding energy, the latter of which are usually interpreted as gravitationally unbound.
This was particularly surprising, as energy-balance arguments suggest that BSs with positive binding energy suffer an instability to fission.
These results do not, however, rule out the possibility of a non-radial unstable mode, or another instability which does not manifest in spherical symmetry.
Such a situation would not be entirely surprising.
Indeed, rotating BSs have been found to suffer non-axisymmetric instabilities \cite{Di_Giovanni_2020}. 
Unlike radial-mode instabilities, these do not generically emerge at extrema of the BS family's mass-radius curves \cite{Siemonsen_2021}.
One might suspect, then, that positive binding energy on a radially stable branch of a BS family is a clue pointing towards a lurking non-radial instability.
In Ref.~\cite{Marks_2025_CP}, we also conjectured that such models may not be possible to form via the gravitational collapse of a diffuse cloud of scalar matter.

In this work, we will address the issues of stability beyond spherical symmetry and dynamical formation.
First, we supplement our earlier perturbed evolutions, which enforced spherical symmetry, with full nonlinear $3+1$ evolutions without dimensional reduction.
To our spherically symmetric BSs we ``add-by-hand'' explicit axisymmetric and non-axisymmetric perturbations, aiming to determine whether these can seed a dynamical instability.
We will focus particularly on the perturbatively stable models with positive binding energy identified in Ref.~\cite{Marks_2025_CP}.
For these, we will show that generic perturbations are still incapable of seeding an instability to fission.
Furthermore, even with perturbations that introduce significant asphericity, the spherical symmetry of the configuration is rapidly restored.

After this, we will turn our attention to the issue of forming BSs from a more diffuse cloud of scalar matter.
For sufficiently massive clouds, we will see that solitonic BSs form as expected, and that the gravitational cooling driving them to equilibrium is generally much faster than for BSs comprised of a scalar field without self-interactions.
When we turn to less massive clouds, we will find evidence in support of our earlier conjecture: stable BSs with positive binding energy do not generically form.
However, we can still observe the formation of compact objects in such cases--- albeit not stationary ones.
Instead, we find that the end-state is a bound configuration that undergoes continual undamped oscillations, reminiscent of the so-called \textit{oscillatons} \cite{Seidel:1991zh} that consist of a self-gravitating massive real scalar field.

The picture that emerges from our results can be summarized by the following main points,
\begin{enumerate}
    \item We find no evidence that positive binding energy necessarily signifies an aspherical and/or nonlinear instability in spherically symmetric solitonic BSs. The predictions of radial perturbation theory are always sufficient.
    \item Nonetheless, perturbatively stable BSs with positive binding energy appear to be strongly disfavoured as outcomes of formation from a diffuse scalar cloud--- we find no cases in which such BSs form. Where solitonic BSs do form, their gravitational cooling is much faster than that of mini BSs.
    \item In some cases, non-equilibrium configurations of scalar matter which undergo large, long-lasting radial oscillations can form instead of equilibrium BSs. 
    
\end{enumerate}

We employ Planck units throughout setting $c=G=\hbar=1$, but keep the
scalar-field mass parameter $\mu$ under whose rescaling our BS solutions respect an appropriate scale symmetry.

\section{Theory and Computational Framework}
\subsection{Solitonic Boson Stars}
First, we briefly review our matter model. Our action consists of a complex scalar field $\varphi$ minimally coupled to gravity,
\begin{eqnarray}
  S=\int  \sqrt{-g} \left\{
  \frac{R}{16\pi}-\frac{1}{2}\left[
  g^{\mu\nu}\nabla_{\mu}\bar{\varphi}\,\nabla_{\nu}\varphi
  +V(\varphi)
  \right]
  \right\}\du^4 x,\quad
  V(\varphi) = \mu^2 |\varphi|^2 \left(
  1-2\frac{|\varphi|^2}{\sigma_0^2}
  \right)^2,
  \label{eq:action}
\end{eqnarray}
where $\sigma_0$ is a constant. Varying $S$ leads to the Einstein-Klein-Gordon equations
\begin{eqnarray}
  && G_{\alpha\beta}=8\pi \,T_{\alpha\beta}\,,
  ~~~~~
  \nabla^{\mu}\nabla_{\mu}\varphi = \frac{\du V}{\du \bar{\varphi}}, \quad
   T_{\mu\nu}
  =
  \frac{1}{2} \nabla_{(\mu} \bar{\varphi}\,\nabla_{\nu)}\varphi
  -\frac{g_{\mu\nu}}{2}
  \left[
  g^{\alpha\beta}\nabla_{\alpha}\bar{\varphi}\,\nabla_{\beta}\varphi
  + V(\varphi)
  \right],
\end{eqnarray}
which we solve numerically with a harmonic ansatz of the form $\varphi(t,r)=A(r)e^{\iu \omega t},$ fixing the central field amplitude $A_0 = A(0)$ and determining the BS frequency $\omega$ via a two-way shooting algorithm; cf. Ref.~\cite{Evstafyeva:2023kfg}.
This results in a one-parameter family of asymptotically regular BS solutions.
Along with the ADM mass $M$, there is a Noether charge $N$ associated with the $U(1)$ symmetry of the complex scalar field,

\begin{equation} \label{eq:noether}
    N = \int \du ^ 3 x \;\sqrt{-g} \;J^0, \quad
\end{equation}
where $J^\mu = \frac{\iu}{2}(\varphi\nabla^\mu\bar\varphi - \bar\varphi\nabla^\mu \varphi)$ is the conserved current.
The \textit{binding energy} is then defined as 
\begin{equation} 
E_B = M - \mu N. 
\end{equation}
Since $N$ can be interpreted as a particle number, this measures the excess energy of the system relative to a configuration in which all scalar matter has been dispersed  to infinity (with zero residual kinetic energy).
Solutions with negative binding energy are thus interpreted as gravitationally bound, and those with positive binding energy as gravitationally unbound.

The parameter space for solitonic BSs with $\sigma_0 \lesssim 0.36$ contains two (radially) linearly stable branches \cite{Santos:2024vdm}.
The first extends from the Minkowski limit $A_0 \rightarrow 0$ to the first local maximum in the curve $M(A_0)$, while the second extends from the first local minimum to the second local maximum.
These are sometimes referred to as the \textit{Newtonian} and \textit{relativistic} branches,  as the second houses models which are much more compact, and therefore constitutes a stronger-gravity regime.

\begin{figure}[t!]
    \includegraphics[width=\linewidth]{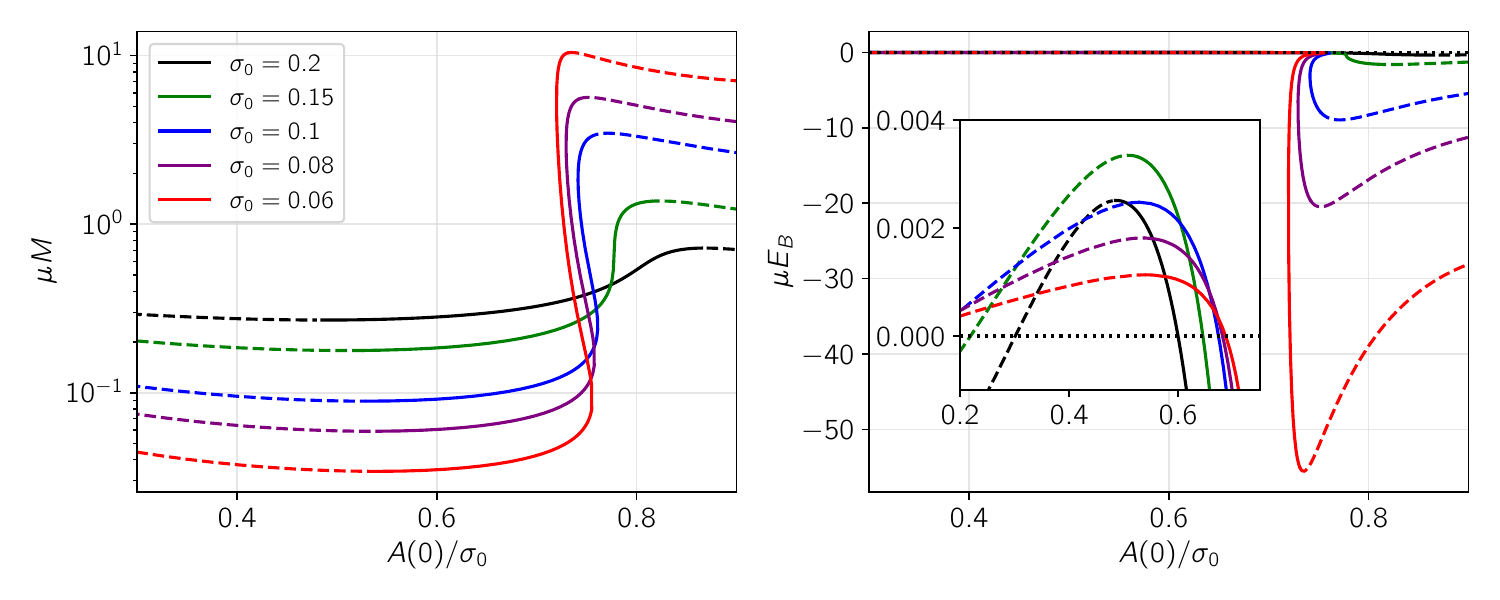}
    \caption{
    The ADM mass $M$ (left) and binding energy $E_B$ (right) against the central scalar-field amplitude
    $A_0$ for families of BS solutions with
    varying parameter $\sigma_0$. Solid lines indicate linearly stable branches, while dashed lines indicate linearly unstable branches--- see Ref.~\cite{Marks_2025} for details of the perturbative calculations. The inset draws attention to the region in which linearly stable models with positive binding energy exist.
    \label{fig:M_and_E}}
\end{figure}

In Fig.~\ref{fig:M_and_E} we plot the mass and binding energy against the central field amplitude for some solitonic families. For $\sigma_0 \lesssim 0.085$ these contain UCOs on the perturbatively stable branch.
The BSs shown in the inset will be of particular interest to us.
Notice that for each family shown, there exist perturbatively stable models with positive binding energy.

\subsection{Evolutions in Spherical Symmetry} \label{sec:spherical_symmetry}
To perform dynamical evolutions in spherical symmetry, we employ the code {\sc
sphericalbsevolver} ({\sc SBSE}) \cite{Marks_Spherical_BS_Evolver}, which uses dimensional reductions of the Baumgarte-Shapiro-Shibata-Nakamura (BSSN) \cite{Baumgarte:1998te,Shibata:1995we} and conformal covariant Z4 (CCZ4) \cite{Alic:2011gg} formulations commonly used in numerical relativity in an arbitrary number of background spacetime dimensions. 
Dimensional reductions  in {\sc SBSE} are based on the so-called \textit{modified cartoon} approach \cite{Pretorius_2004, Cook:2016soy}, where all physical quantities are projected onto a particular coordinate axis, which we take to be the $z$-axis. Partial derivatives in off-axis directions can then be replaced via the tensor transformation laws. Here we specialize to $D = 4$ spacetime dimensions so that we have an $SO(3)$ symmetry, and take the spatial coordinates on a given time slice to be
\begin{equation}
x^i = (z, w^1, w^2).
\end{equation}
Furthermore, we will use early Latin indices $a,b,c,...$ to run over the two off-axis coordinates, so that we can also write the above as $(z, w^a)$. 
Where any of the last three coordinates could be used interchangeably, we will simply write $w$.
The metric we use for dynamical evolutions is then given by
\begin{equation}
    \du s^2 = -\alpha^2 \du t^2 + \gamma_{zz}(\du z^2 + 2 \beta^z \du z \du t) + \gamma_{ab}\du w^a \du w^b,
\end{equation} 
where $\alpha$ is the lapse, $\beta^i = (\beta^z, 0, 0)$ is the shift vector, and  $\gamma_{ij} = \mathrm{diag}(\gamma_{zz}, \gamma_{ww}, \gamma_{ww}) $ is the metric of the spacelike hypersurface.

We apply this dimensional reduction to the BSSN and CCZ4 systems, which introduce the following evolution variables,
\begin{itemize}
\item The conformal factor $\chi = (\det\gamma)^{-\frac{1}{3}},$
\item The conformally rescaled metric $\cgamma_{ij} = \chi\gamma_{ij},$
\item The mean curvature $K = \gamma^{ij}K_{ij},$
\item The conformally rescaled traceless extrinsic curvature $\tilde{A}_{ij} = \chi\left(K_{ij} - \frac{1}{3} K\gamma_{ij}\right)$, and finally
\item The contracted conformal Christoffel symbols, $\tilde{\Gamma}^i = \frac{1}{2}\cgamma^{il}\cgamma^{jk}\left(\partial_j \cgamma_{lk} + \partial_k \cgamma_{lj} - \partial_l \cgamma_{jk} \right). $
\end{itemize}
A complete description of the evolution systems used by {\sc SBSE} in isolation can be found in Appendix A of Ref.~\cite{Marks_2025_B}. 
In this work, we use the CCZ4 system for the stability runs presented in Sec.~\ref{sec:spherical_results}, with damping parameters $\kappa_1 = 0.1 / M,$ $\kappa_2 = 0,$ $\kappa_3 = 1$ (cf. Ref.~\cite{Alic:2011gg}).
For the formation runs in Sec.~\ref{sec:formation_results}, we use the BSSN system as the relevant dynamical length-scales can vary significantly over the course of an evolution, and because these runs are shorter the long-term growth of constraint violations in less of a concern.

To reduce the early-time gauge dynamics, we always transform our initial data to isotropic coordinates so that the metric has the form
\begin{equation}
    \du s^2 = -\alpha^2\du t^2 + \psi^4\left(\du r^2 + r^2\du \Omega^2 \right);
\end{equation}
see Ref.~\cite{Helfer_2022} for more details.

 {\sc SBSE} uses fourth-order discretization in both space and time, via the method of lines with a Runge-Kutta integrator for the time step.
It also supports arbitrary fixed mesh refinement.
Unless otherwise indicated, all runs in spherical symmetry are performed with spatial resolution $\mu\Delta z = 1/16$ on the finest level, with a Courant factor $\mathcal{C} := \Delta t / \Delta z = 0.4.$
We use outgoing Sommerfeld boundary conditions at the outer boundary for all numerical fields,
\begin{equation}
    \partial_t \phi + \partial_r \phi +\phi - \phi_\infty = 0
\end{equation}
where $\phi$ can represent any grid variable with asymptotic value $\phi_\infty$.

To determine whether a black hole has dynamically formed, we check at every timestep for the presence of an apparent horizon. 
In our formalism, this reduces to a search for solutions to the equation,
\begin{equation}
    \sqrt{\frac{\chi}{\cgamma_{zz}}}\left[\frac{2}{z} + \frac{1}{\cgamma_{ww}}\left(\partial_z \cgamma_{ww} - \cgamma_{ww}\frac{\partial_z\chi} {\chi} \right) - \frac{2}{\cgamma_{ww}} \left(\tilde{A}_{ww} + \frac{1}{3}K\cgamma_{ww} \right) \right] = 0.
\end{equation}

\subsection{Full 3+1 Evolutions} \label{sec:full_3+1}

We perform fully nonlinear dynamical evolutions with no enforced symmetry using {\sc grchombo} \cite{Andrade:2021rbd, Radia:2021smk}, an open-source code capable of evolving the BSSN and CCZ4 formulations of numerical relativity with full adaptive mesh refinement (AMR). 
For the matter sector, we use the {\sc exozvezda} extension, described in Refs.~\cite{Helfer_2022,Evstafyeva:2023kfg,Croft:2022bxq, Croft:2022gks}, to handle the evolution of the scalar field and compute related diagnostic quantities.
We primarily use the CCZ4 formulation due to its constraint-damping properties. 
Previous work using CCZ4 for evolutions of isolated BSs has found excellent agreement with codes based upon a generalized harmonic gauge formalism \cite{Evstafyeva:2025mvx}. 

Unless otherwise indicated, each 3+1 run presented has been performed on a grid of uniform edge length $L =1024 M$ 
with 6 AMR levels. The grid spacing on the finest level is $\mu\Delta x = 1/12$.
The tagging criterion for mesh refinement is based on second derivatives of the conformal factor, scalar field, and energy density; see Section 3.5 of Ref.~\cite{Radia:2021smk} for more details.

In Appendix \ref{sec:convergence} we present some resolution studies demonstrating the numerical convergence of our results, for evolutions both with and without enforced spherical symmetry.

\section{Nonlinear Stability Results } \label{sec:nonlinear_stability}
\subsection{Spherical Symmetry} \label{sec:spherical_results}
First, we briefly review and extend the results obtained by evolving perturbed BSs in enforced spherical symmetry.
 In this section, we prescribe perturbations using a Gaussian profile of the form 
 \begin{equation}
     \delta \varphi = a \exp\left(-(r - r_0)^2 / k^2\right), \quad \delta \Pi_\varphi = -\iu\omega \delta\varphi / \alpha , \label{eq:spherical_pert}
 \end{equation}
 for real constants $a, r_0, k$ where $\alpha$ is the lapse.
 Following the arguments of Ref.~\cite{Alcubierre_2019}, this prescription for the field momentum ensures that the Noether charge is conserved at leading order in the perturbation size.
 Once perturbations are added, we re-solve the Hamiltonian constraint and polar slicing condition, while the momentum constraint remains automatically satisfied.

We list the dynamical fates alongside some diagnostic quantities for a series of runs in Table \ref{tab:tab1}. 
These are divided into three categories.
First, there are general examples of perturbatively stable and unstable models located close to the ends of their respective stability branches (prefixed A).
Second, there are linearly stable models with $E_B > 0$ (prefixed B).
Finally, there are ultracompact models (prefixed C). 
All stable evolutions have been run for a time of at least $\mu t = 10^5$ with no sign of instability. 
Note that the runs \texttt{A1}, \texttt{A2}, \texttt{A3}, \texttt{B1}, \texttt{B2}, \texttt{B3}, \texttt{C1}, \texttt{C2}, \texttt{C3} were included previously in Ref.~\cite{Marks_2025_CP}.
We supplement them here with additional evolutions using BS models whose behaviour we will explore in more detail in the following sections.

\begin{table}[t]
\centering
\caption{Results of our time evolutions in spherical symmetry, showing the central amplitude $A_0$, solitonic parameter $\sigma_0$, BS mass $M$, binding energy $E_B$, compactness $C_{99} = M / r_{99}$ with $r_{99}$ defined as the radius enclosing 99\% of the mass, perturbation parameters $a$, $k$, $r_0$, whether the model is linearly stable, and the evolution outcome.}
\label{tab:tab1}
\begin{tabular}{@{}ccccccccccc@{}}
\toprule
\textbf{Label} &$A_0$ &$\sigma_0$  &$\mu M$  & $\mu E_B$  &$C_{99}$   &$a$  &$\mu k$  &$\mu r_0$  & Lin. Stable?  & Outcome  \\ \midrule

 \texttt{A1}&0.09  &0.2   &0.271  &0.00236  &0.0403  &0.001  &0.1  &10.0  &No  &Dispersion  \\ 
 \texttt{A2}&0.17  &0.2  &0.713  &-0.334  &0.179  &0.005  &1.0  &3.0  &Yes  &Stable  \\ 
 \texttt{A3}&0.18  &0.2  &0.706  &-0.325  &0.190  &-0.001  &1.0 & 3.0  &No  &BH Collapse \\ 
 \texttt{A4}&0.12  &0.15  &1.321  &-1.481  &0.228  &0.0005  &1.0 & 10.0  &Yes  &Stable  \\
 \texttt{A5}&0.125  &0.15  &1.368  &-1.199  &0.2482  &0.001  &1.0 & 1.0  &No  & BH Collapse  \\
 \texttt{A6}&0.125  &0.15  &1.368  &-1.199  &0.2482  &-0.001  &1.0 & 1.0  &No  &  Migration  \\
 \texttt{A7}&0.09  &0.2   &0.271  &0.00236  &0.0403  &-0.001  &0.1  &10.0  &No  &Dispersion  \\ 
 \texttt{B1}&0.035  &0.06  &0.0345  &0.00109  &0.00516  &-0.001  &1.0  &2.0  &Yes  &Stable  \\ 
 \texttt{B2}&0.045  &0.08  &0.0593  &0.00179  &0.00867  &0.001  &1.0  &3.0  &Yes  &Stable  \\ 
 \texttt{B3}&0.055  &0.1  &0.0893  &0.0245  &0.0131  &-0.001  &2.0  &0.0  &Yes  &Stable  \\ 
 \texttt{B4}&0.085  &0.15  &0.180  &0.00296  &0.0293  &0.0002  &1.0  &20.0  &Yes  &Stable  \\ 
 \texttt{B5}&0.09  &0.15  &0.184  &0.00222  &0.0315  &-0.002  &1.0  &0.0  &Yes  &Stable  \\ 
 \texttt{B6}&0.1  &0.2  &0.271  &0.00250  &0.0435  &0.0001  &1.0  &5.0  &Yes  &Stable  \\
 \texttt{C1}&0.044  &0.06  &10.43  &-55.9  &0.328  &0.001  &0.5  &10.0  &Yes  &Stable  \\ 
 \texttt{C2}&0.045  &0.06  &10.39  &-52.2  &0.340  &-0.001  &1.0  &5.0  &No  &BH Collapse  \\ 
 \texttt{C3}&0.06  &0.08  &5.65  &-20.5  &0.313  &-0.0005  &0.5  &10.0  &Yes  &Stable  \\
 \texttt{C4}&0.06  &0.08  &5.65  &-20.5  &0.313  &0.0005  &0.5  &10.0  &Yes  &Stable  \\ 
 \texttt{C5}&0.045  &0.06  &0.06  &-52.2  &0.340  &0.001  &1.0  &5.0  &No  &BH Collapse  \\
 \texttt{C6}&0.0602  &0.08  &5.66  &-20.6  &0.315  &0.0005  &0.5  &10.0  &No  &BH Collapse  \\ 
 \texttt{C7}&0.0602  &0.08  &5.66  &-20.6  &0.315  &-0.0005  &0.5  &10.0  &No  &BH Collapse  \\ 
\bottomrule
\end{tabular}
\end{table}

As expected from previous work, the explicit perturbations added here introduce no departures from the predictions of  linear theory. Linearly unstable models either collapse to BHs, migrate (if $E_B < 0$) or disperse (if $E_B > 0$).
Likewise, linearly stable models remain stable in perturbed evolutions so long as the perturbation does not change the total Noether charge so much that no BS model with that value exists on the perturbatively stable branch.
This latter condition is particularly relevant for the linearly stable models with $E_B > 0$, which are the lowest-charge models on the relativistic branch, and for ultracompact objects, which are the highest-charge.
Nevertheless, even with perturbations that remove scalar matter and hence further increase $E_B$ (cf. inset of Fig.~\ref{fig:M_and_E}), there is no sign of a dynamical instability for runs in the B-group.
Similarly, even for runs that add scalar matter, we see no dynamical instability in the C-group.

We also draw attention here to a few interesting pairs of runs. 
The pairs $\{\texttt{A5}, \texttt{A6}\}$, $\{\texttt{A1}, \texttt{A7}\}$, $\{\texttt{C2}, \texttt{C5}\}$ and $\{\texttt{C6}, \texttt{C7}\}$ all evolve the same unstable BSs with perturbations that differ only by sign.
That is, one adds scalar matter, and the other removes it.
In the case $\{\texttt{A5}, \texttt{A6}\}$, this change alters the dynamical fate of the BS: the BS to which Noether charge was added collapses to a BH, while the other migrates back to the stable branch.
However, the pair $\{\texttt{A1}, \texttt{A7}\}$ corresponding to two highly diffuse BSs disperse, while the pairs of ultracompact BSs $\{\texttt{C2}, \texttt{C5}\}$ and $\{\texttt{C6}, \texttt{C7}\}$ suffer collapse to BHs in all cases.
This is indicative of a more general pattern seen in the dynamical fates of perturbed unstable BSs.
Very diffuse models disperse irrespective of the character of the perturbation, while very compact models always collapse to BHs.
Only in the intermediate region can the character of the perturbation affect the dynamical fate of the BS--- typically, changing it from migration to BH collapse.
We remark that this may have interesting ramifications for the kinds of critical phenomena that can be observed using BS models, which so far have received attention only in the non-interacting case \cite{Hawley_2000, Lai_2004, Lai_2007, Jimenez_2022}.

Overall, our earlier findings in spherical symmetry are reinforced: there are no departures from the linear predictions. One can contrast this with the case of neutron star models, where nonlinearities in the radial oscillations have been found in numerical studies to affect stability to spherical dynamics \cite{Sperhake_2001n}.



\subsection{Beyond Spherical Symmetry} \label{sec:beyond_spherical}
We now turn our attention to full $3+1$ evolutions using {\sc grchombo}.
Here, we will be particularly interested in probing the radially stable models with positive binding energy.
We note that the issue of nonlinear stability of solitonic BSs has been previously addressed numerically in Refs.~\cite{Ge:2024itl,Marks_2025}.
However, Ref.~\cite{Ge:2024itl} used BS families with a large enough value of the solitonic parameter $\sigma_0$ that no perturbatively stable BSs with positive binding energy exist, and furthermore made use of a code that enforces axisymmetry.
Ref.~\cite{Marks_2025}, on the other hand, was focused on the nonlinear stability of UCOs, which are strongly gravitationally bound.
The stability of our ``gravitationally unbound'' yet radially stable BSs has thus not yet been subjected to numerical study in a full 3+1 context.
We will focus here on two models, whose properties are given in Table~\ref{tab:tab2}.
Of the perturbatively stable models with $E_B > 0$ that we have constructed, \texttt{S15} is close to maximizing the binding energy in absolute terms, while \texttt{S06} is close to maximizing the ratio $E_B / N$. 
Thus, if positive binding energy signifies a non-radial instability, we would expect it to grow most quickly for one or both of these models.

\begin{table}[t]
\centering
\caption{Properties of the two perturbatively stable BSs with positive binding energy studied in this section, including the solitonic parameter $\sigma_0$, central amplitude $A_0$, mass $M$, binding energy $E_B$, and compactness $C_{99}$.}
\label{tab:tab2}
\begin{tabular}{@{}cccccc@{}}
\toprule
\textbf{Label}  &$\sigma_0$ &$A_0$  &$\mu M$  & $\mu E_B$  &$C_{99}$   \\ \midrule

 \texttt{S15} &0.15  &0.085  &0.180  &0.00296  &0.0293   \\ 
 \texttt{S06} &0.06 &0.035 &0.0345  &0.00109  &0.00516    \\ 
\bottomrule
\end{tabular}
\end{table}

To study the non-radial dynamics, we wish to introduce explicit aspherical perturbations to our BSs. We achieve this by modifying the scalar field profile according to,
\begin{equation}
\varphi \rightarrow e^{-2 \delta\cos(m\phi)}\varphi, \quad \Pi_\varphi \rightarrow \frac{\omega}\alpha e^{-2 \delta \cos(m\phi)}\Pi_\varphi, \label{aspherical_pert}
\end{equation}
where $\delta$ represents the perturbation magnitude, $\phi$ is the azimuthal angle with respect to some pair of orthogonal reference axes, and $m$ is an integer specifying the azimuthal index of the perturbation.
Of course, since our unperturbed BSs are spherically symmetric, a symmetry axis remains once this perturbation is applied.
To study the effect of seed profiles without an $SO(2)$ symmetry, then, we will also perform runs using two distinct perturbations of the form given by Eq.~\eqref{aspherical_pert} prepared in orthogonal planes.
We will use $\epsilon$ and $n$ to denote the magnitude and azimuthal index of these additional perturbations.

To track the evolution of deviations from spherical symmetry, we follow Ref.~\cite{Evstafyeva:2025mvx} and introduce an azimuthal decomposition of the complex field amplitude,
\begin{equation} \label{eq:phi_modes}
    \Phi_j := \frac{1}{\max |\varphi|} \int \du^3x \sqrt{\gamma}  |\varphi|e^{\iu j\phi},
\end{equation}
so that $\Phi_j$ encodes the power contained in the $j$th mode.
We track the evolution of $\Phi_j$ up to $j = 10$.
To account for a small numerical drift in the position of the BS, we perform a Gaussian fit on each coarse timestep to track the BS location, and use the center extracted in this manner as the coordinate origin for the purposes of Eq.~\eqref{eq:phi_modes}.

\begin{figure}[t]
    \includegraphics[width=\linewidth]{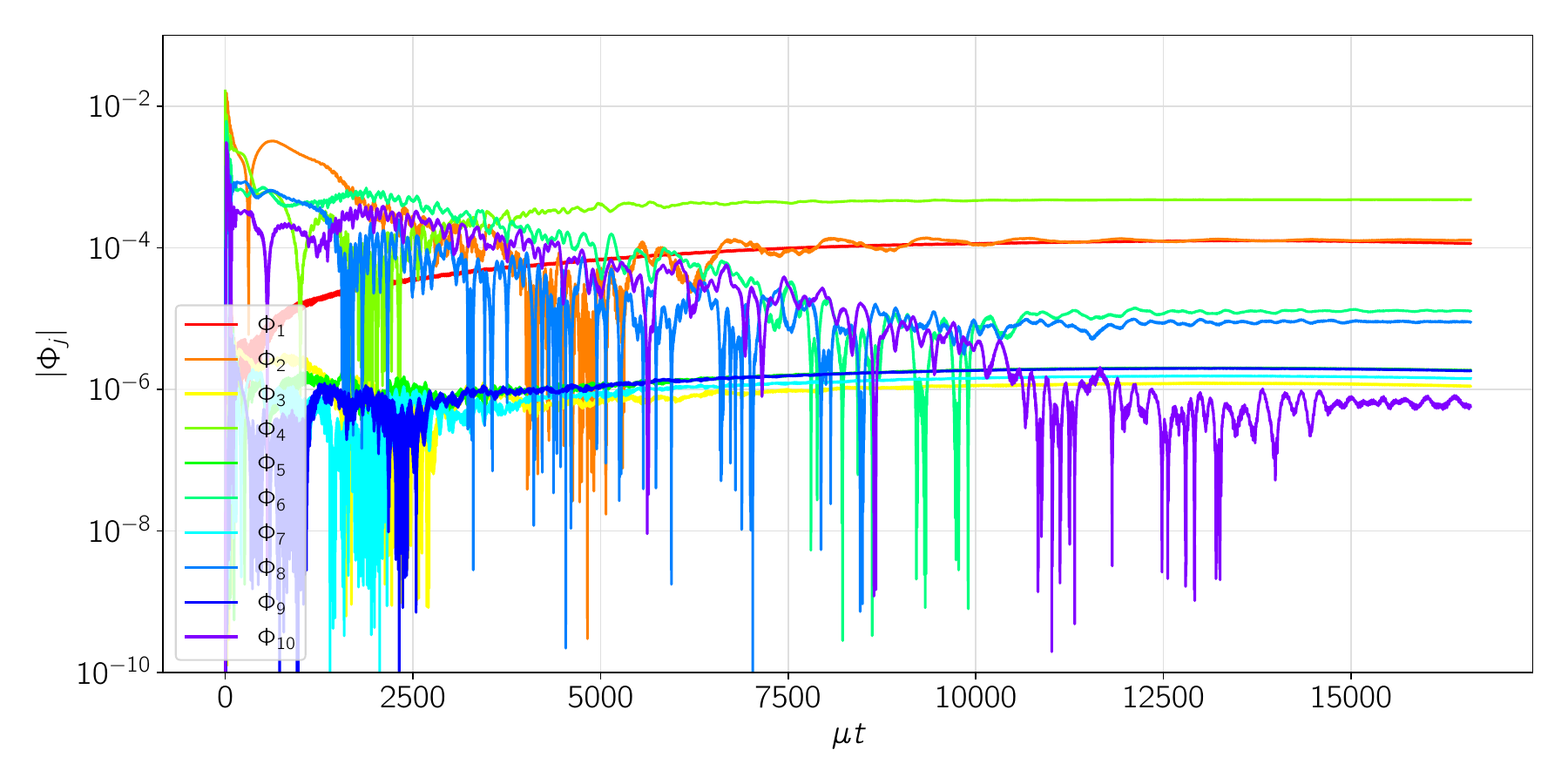}
    \caption{The magnitudes of the modes $\Phi_1$ through $\Phi_{10}$ in the azimuthal decomposition given by Eq.~\eqref{eq:phi_modes} over time for evolution \texttt{IA}. Note that the $m  = 4$ mode, the largest at late times, is the one we have explicitly excited.
    \label{fig:IA_modes}} 
\end{figure}

We show an example of the time evolution of the azimuthal decomposition given by Eq.~\eqref{eq:phi_modes} in Fig.~\ref{fig:IA_modes}.
Notice that at early times, the $m = 4$ mode, which we have manually excited, is dominant (along with $m = 2$).
However, this initial asphericity rapidly decays by almost two orders of magnitude, before these and all other modes saturate at finite values, showing no signs of unbounded growth.
That there should be some residual asphericity is unsurprising, as our imperfect Sommerfeld boundary conditions introduce small, continual perturbations throughout the evolution.
Interestingly, there seems to be an exchange of power between the $m = 1$ and $m = 2$ modes, with the two saturating at nearly identical values.

We list the 3+1 evolutions performed in Table~\ref{tab:tab3}.
In all cases, there is no clear sign of instability on the timescale for which we evolve.
The Noether charge measured in the computational domain remains conserved to within $0.01 \%$, and the value of $\max |\varphi|$ begins to oscillate at some fixed frequency.
\begin{table}[t]
\centering
\footnotesize
\caption{List of all $3+1$ evolutions performed involving aspherically perturbed BSs, showing the BS evolved, perturbation parameters $m$, $\delta$, $n$, $\epsilon$, and the coordinate time $T$ for which the model was evolved with no sign of instability.}
\label{tab:tab3}
\begin{tabular}{@{}ccccccc@{}}
\toprule
\textbf{Label}  & Model  &$m$ &$\delta$  &$n$  & $\epsilon$  &$\mu T$   \\ \midrule

\texttt{IA} &  \texttt{S15}  &4 &0.01  &--- & --- & $1.6 \times 10^{4} $   \\ 
\texttt{IB} &  \texttt{S15}  &3 &0.05  &--- & --- & $5 \times 10^{3} $   \\
\texttt{IC} &  \texttt{S15}  &1 &0.01  &8 & 0.01 & $5 \times 10^{3} $   \\
\texttt{IIA} &  \texttt{S06}  &6 &0.01  &--- & --- & $1.2 \times 10^{4} $   \\ 
\texttt{IIB} &  \texttt{S06}  &2 &0.05  &--- & --- & $5 \times 10^{3} $   \\ 
\texttt{IIC} &  \texttt{S06}  &3 &0.01  &4 & 0.01& $5 \times 10^{3} $   \\ 

\bottomrule
\end{tabular}
\end{table}

As is shown in Fig.~\ref{fig:IA_modes}, the manually excited mode, there $m = 4$, always decays rapidly at early times.
We note that this strengthens the observation made in other contexts that BSs tend to be exceptionally efficient at converting non-radial excitations to a purely radial oscillation.
For instance, Ref.~\cite{Siemonsen_2023} shows that rotating BSs do not generically form as the result of binary inspirals, even if the initial configuration has sufficient mass and angular momentum to support them.
Instead, the generic outcome is a spherically symmetric ground-state BS, with all angular momentum being lost in scalar and gravitational radiation.

Ultimately, our results yield no evidence of a non-radial instability in either model, linear or otherwise.
We conclude that these BSs, despite having positive binding energy, are legitimately stable under small perturbations once formed.

\section{Formation Results} \label{sec:formation_results}
We now turn our attention to the study of compact object formation via the collapse of non-equilibrium clouds of scalar matter with a solitonic potential.
All runs in this section are performed in spherical symmetry, using {\sc SBSE}.
For initial data, we use a simple Gaussian cloud of the form
\begin{equation}
    \varphi = Ae ^{-r^2 /s^2},  \quad \Pi_\varphi = -\frac{\iu \Omega}{2\alpha}\varphi \label{eq:scalar_cloud}
\end{equation}
where $A,$ $s,$ and $\Omega$ are real constants, so that our initial data roughly mimics the structure of a BS by oscillating instantaneously at a uniform frequency. 
Unless otherwise indicated, we will take $\Omega = 1$ from this point onwards.
Assuming a moment-of-time symmetry so that $K_{ij} = 0$, the momentum constraint is automatically satisfied on the initial slice.
We then solve the Hamiltonian constraint and polar slicing condition via quadrature in our initial isotropic coordinates before evolving.

For a given $\sigma_0$, all BSs on the relativistic branch have Noether charge $N$ above some minimum value, achieved by the marginally stable model at the first minimum of the $A_0$-$N$ curve: call this value $N_0$.
The Noether charge of such BSs with $E_B < 0$ achieves a similar, larger minimum, which we call $N_1$.
In general, we will see that the Noether charge $N$ of our initial Gaussian configuration relative to $N_0$ and $N_1$  significantly affects the possible outcomes regarding compact object formation.
Of course, for $N < N_0$ it is no surprise that formation of a BS on the relativistic branch is impossible.
The salient question is whether $N < N_1$ is also consistently required.

To identify the BS formed from dynamical evolution, it is usually sufficient to determine the limiting value of the central field amplitude $A_0(t)$ at late times. 
The frequency $\omega,$ mass $M$ and Noether charge $N$ then agree well with those predicted by computing an equilibrium solution \textit{a priori}.
However, as $\sigma_0$ becomes small, a region of the parameter space emerges in which the curve $M(A_0)$ becomes multivalued. 
In this region the equilibrium BS model is very sensitive to small variations in the central amplitude, and so naive use of the aforementioned method leads to large uncertainties.
Therefore, we instead extract the Noether charge at late times, once essentially all the unbound scalar matter has had a chance to leave the computational domain.

\begin{figure}[t]
    \includegraphics[width=\linewidth]{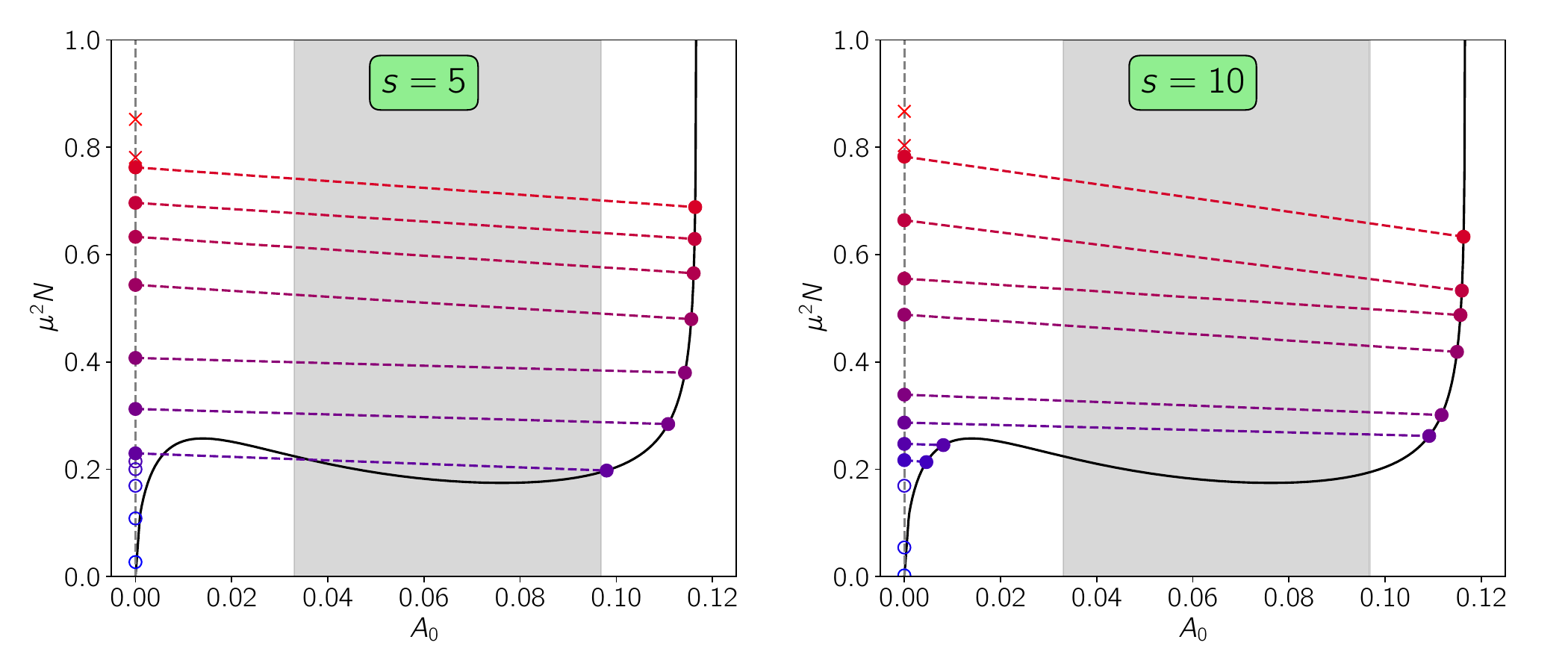}
    \caption{Each point on the vertical line at $A_0 = 0$ shows the Noether charge of the initial scalar cloud \eqref{eq:scalar_cloud} for a range of maximum cloud amplitudes $A$ with $\sigma_0 = 0.15$, for $s = 5$ (left) and $s = 10$ (right). The outcome of dynamical evolution is indicated as follows. Dispersion of all scalar matter is represented by an empty circle. Formation of a BH is represented by a cross. Finally, if a BS forms, its position on the Noether charge-central amplitude curve is shown with a solid dot, and connected to the corresponding initial Noether charge by a dashed line. The shaded region of the parameter space has positive binding energy.
    \label{fig:formation_s015}} 
\end{figure}

In Fig.~\ref{fig:formation_s015}, we graphically illustrate the end-state of the initial scalar cloud's collapse (or dispersion) for a selection of evolutions with $\sigma_0 = 0.15$. 
We compare two values of the initial cloud extent $s$.
Some features are common: in particular, notice neither case provides an example of the formation of a BS with positive binding energy.
With $s = 5$, BSs on the first stable branch never form; instead, those clouds with insufficient Noether charge to form a gravitationally bound BS on the second stable branch (i.e. those with $N \le N_1$)  simply disperse.
With $s = 10$, however, BSs on the first stable branch do form, including for clouds with $N > N_1$.
This result is not surprising when we recall that the first stable branch is significantly more diffuse, with compactness $C_{99}$ of less than $0.013$ throughout, while on the second stable branch it is at least $0.028$.
In all cases of BS formation, we see from the decrease in $N$ that some scalar matter has been lost relative to the original configuration.
This is due to the monopole-radiation mechanism of \textit{gravitational cooling}, which is already known to play an important role in the formation of mini BSs \cite{Seidel:1993zk} as well as Proca stars \cite{DiGiovanni:2018bvo}.

\begin{figure}[t!]
    \includegraphics[width=\linewidth]{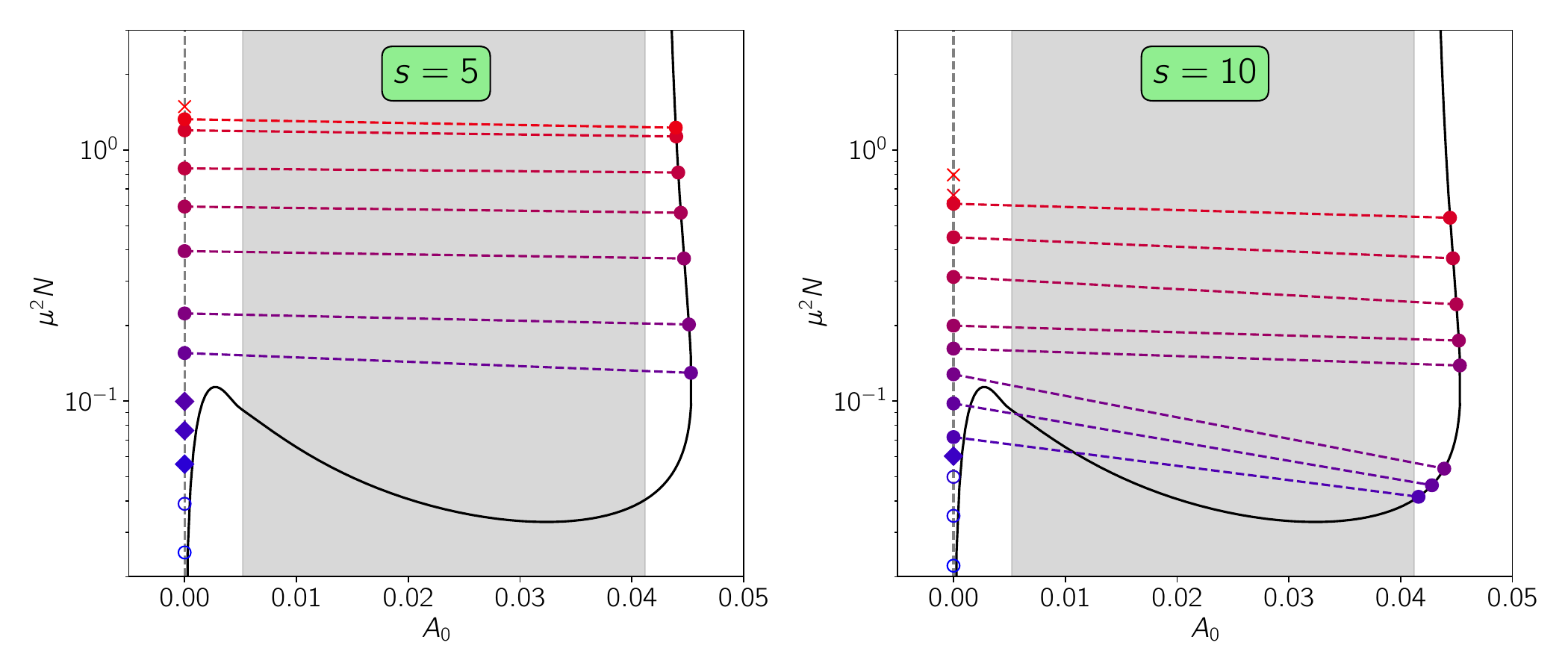}
    \caption{The same as Fig. ~\ref{fig:formation_s015}, but using a solitonic potential with $\sigma_0 = 0.06$. We have also introduced diamonds on the grey vertical line to denote a new dynamical end-state seen in this case: formation of a compact object that undergoes continual radial oscillations, which we call a \textit{pseudo-oscillaton} in the main text. 
    \label{fig:formation_s06}} 
\end{figure}

In Fig.~\ref{fig:formation_s06}, we similarly illustrate the end-states obtained using a potential with $\sigma_0 = 0.06$, whose degenerate vacua are close enough to support the ``thin-shell" configurations capable of yielding perturbatively stable ultracompact objects \cite{Boskovic:2021nfs, Collodel:2022jly, Marks_2025}.
Once again, we find that no BSs with positive binding energy form,
but this time we see no instances of BS formation on the first stable branch.
Instead, there is a new phenomenon: formation of a non-stationary compact object that oscillates for long periods in an undamped manner.
The central amplitude of these objects lies close to the degenerate vacuum value $|\varphi| = \sigma_0 / \sqrt{2}$; however their mass, Noether charge, and complex oscillation frequency are inconsistent with any equilibrium BS model possessing that central amplitude.
Moreover, while the damping of oscillations around an equilibrium BS is not unlimited in practice due to the continual propagation of noise from the outer boundary, the magnitude of these particular oscillations is not reduced when we increase the size of the computational grid, and indeed they persist even on a grid so large that the outer boundary remains out of causal contact with essentially all the scalar matter for the entire evolution.
Because of the analogy with similarly non-stationary spherically symmetric configurations of a massive real scalar field, we will refer to these objects as \textit{pseudo-oscillatons}.

\begin{figure}[t!]
    \includegraphics[width=\linewidth]{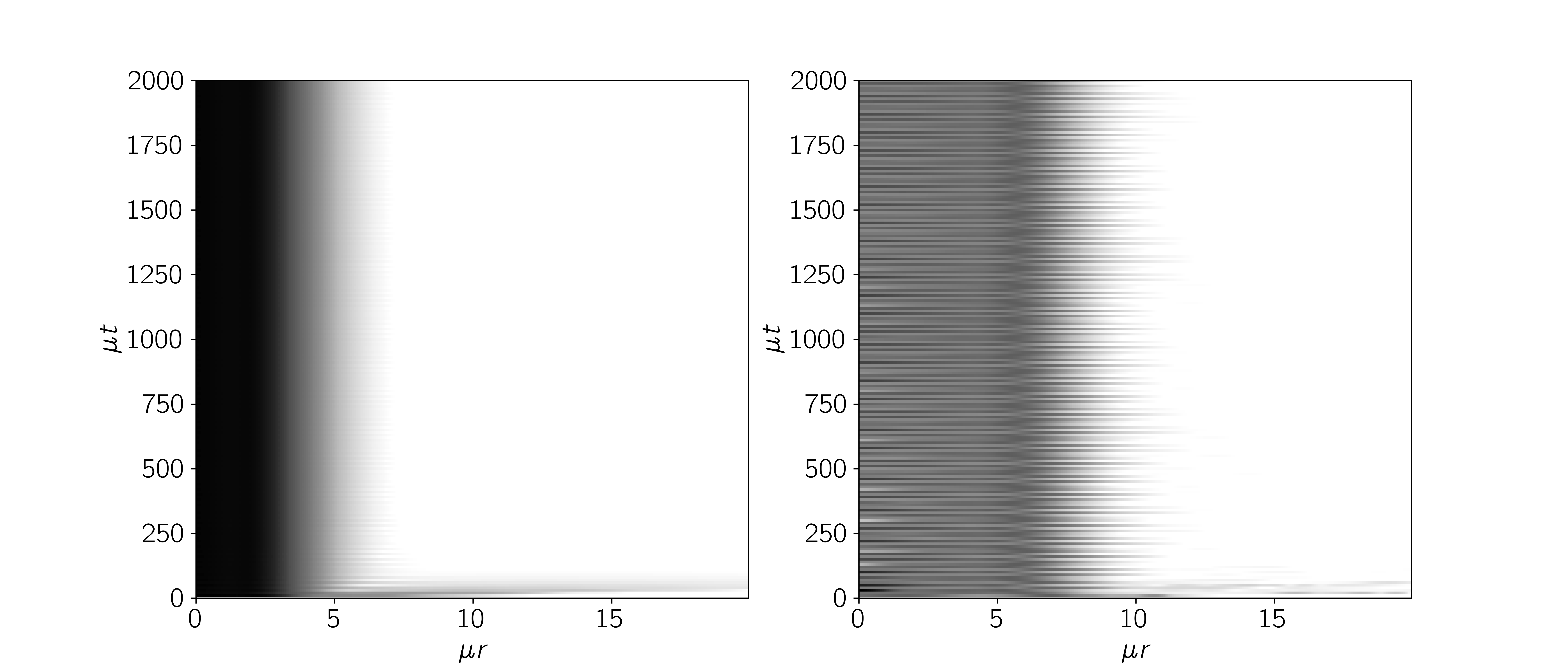}
    \caption{Spacetime diagrams showing radial profiles of the (log-scaled) energy density $\mu^{-2} \rho$, for a configuration forming a boson star with $\sigma_0 = 0.06$, $A = 0.04$, $s = 5$ (left) and one forming a pseudo-oscillaton with $\sigma_0 = 0.06$, $A = 0.02$, $s = 5$ (right). Note that these evolutions were performed on a grid of radius $R = 2000\mu^{-1}$ so that the outer boundary is not in causal contact with the grid center for the duration shown; boundary effects therefore cannot explain the long-lasting oscillations seen on the right.
    \label{fig:formation_rho}} 
\end{figure}

In Fig.~\ref{fig:formation_rho}, we compare a profile of the energy density over time for an evolution resulting in BS formation to one resulting in pseudo-oscillaton formation.
Notice that in both cases, the initial dynamics of each configuration are rapidly damped away on timescales of $\mu t \sim 100$.
This can be contrasted with the mini BS case, where gravitational cooling is extremely slow by comparison \cite{Seidel_Suen_1990, Guzman_2009, Kain:2021rmk}.

\begin{figure}[t!]
    \includegraphics[width=\linewidth]{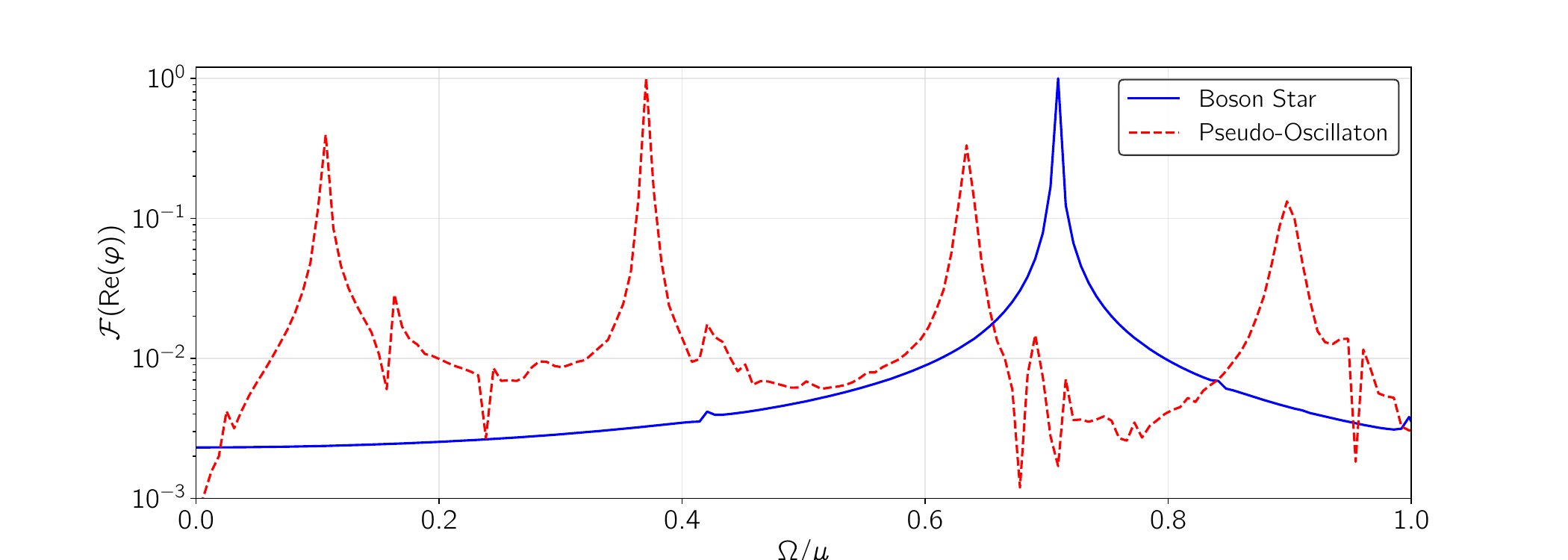}
    \caption{Power spectrum of the central real part of the scalar field $\varphi$ for the evolutions used in Fig.~\ref{fig:formation_rho}, using a time window beginning at $\mu t = 200$ when the initial dynamics have largely settled down.
    \label{fig:formation_fourier}} 
\end{figure}

A possible characterization of these pseudo-oscillatons can be found in the multi-oscillating boson stars constructed by Choptuik, Masachs and Way \cite{Choptuik:2019zji}.
These are compact solutions to the Einstein-Klein-Gordon system whose field profiles are given by quasiperiodic functions.
Thus, a multi-oscillator on $k$ frequencies has the schematic profile,
\begin{equation}
    f(t,r) = \sum_{n_1, ..., n_k}A_{n_1,...,n_k}(r)e^{i(n_1\omega_1 + ... + n_k \omega_k)t},
\end{equation}
where $f$ can represent any observable variable. As evidence for this, in Fig.~\ref{fig:formation_fourier} we show Fourier transforms of the final state for the two evolutions shown in Fig. ~\ref{fig:formation_rho}.
While the power spectrum corresponding to a BS is sharply peaked at a single value--- the BS frequency $\omega$--- in the pseudo-oscillaton case, there are four significant peaks in the range $[0, 1]$.
We therefore conjecture that this object can be construed as a multi-oscillator on up to 4 frequencies.
Demonstrating this would require us to construct multi-oscillator families for the solitonic family, which is likely to be highly involved in the case of small $\sigma_0,$ where even ordinary BSs become difficult to compute by standard methods.
We therefore consider it beyond the scope of this work.

If our pseudo-oscillatons are indeed multi-oscillators for the solitonic family, this would provide, to our knowledge, the first examples of multi-frequency models that are energetically preferred to ordinary single-frequency BSs.
We note in particular that the late-time binding energy, obtained by dynamically extracting the ADM mass and Noether charge, is negative for each pseudo-oscillaton we form.

\section{Conclusions}\label{sec:Conclusions}
We have extended the results of Ref.~\cite{Marks_2025_CP} in two major ways, each helping to clarify the connection between binding energy and stability for solitonic boson stars.
First, we have dynamically evolved radially stable BSs with positive binding energy in full 3+1 numerical relativity under aspherical perturbations, finding no evidence of a non-radial instability.
Our mode analyses show that most of the power contained in non-radial oscillations is rapidly converted to the radial sector.
Thus, the first possibility suggested in our previous work--- that positive binding energy in the radially stable regime suggests a non-radial instability--- now appears unlikely.
A caveat is that an instability with an extremely slow growth rate, or one only manifesting at very large angular indices, may not be evident from our evolutions.
Given that the models in question are highly diffuse, however, we consider the latter possibility unlikely.

Second, we have studied the formation of solitonic BSs from the collapse of spherical clouds of scalar matter.
Here, we find strong evidence for the other conjecture raised in Ref.~\cite{Marks_2025_CP}: that stable solitonic BSs with positive binding energy do not generically form via the usual mechanisms of collapse and gravitational cooling.
Indeed, we find no instances in which models with positive binding energy form.
This is despite the fact that some of our configurations allow the formation of models with $E_B$ very close to zero (but negative). 
Yet tuning the perturbation such that we might expect to form an unbound BS by extrapolation, we invariably find that another end-state occurs instead.
These alternative end-states include formation of a BS on the first stable branch (where the binding energy is strictly negative), dispersion of all scalar matter, and formation of a new object that we have called a \textit{pseudo-oscillaton}.
As a possible explanation for these pseudo-oscillatons, we have proposed the multi-oscillating boson stars described in Ref.~\cite{Choptuik:2019zji}.

Our results may have ramifications beyond the particular matter model used in this work. Other massive-field matter models supporting degenerate vacua, such as the axionic potential \cite{Siemonsen_2021, Arabameri:2026qgc}, are likely to support similarly unbound-but-stable models. 
The fact that compact objects can form with particle number even smaller than that required for a bound BS is promising for the prospects of UCO formation, which could in principle arise by the gradual accretion of scalar matter onto an initial pseudo-oscillaton.
Moreover, we have seen that the unbound-but-stable regime is an intriguing starting point for studying the nonlinear stability of multi-oscillators--- a natural question, left unaddressed in Ref.\cite{Choptuik:2019zji}, is whether these solutions will inevitably decay into single-frequency BSs, and our results suggest that in at least some cases they may not.

Finally, we comment on possibilities raised for identifying critical phenomena.
The BSs we have focused on are stable under small perturbations but seemingly disfavored as dynamical end-states.
It therefore seems reasonable that some critical perturbation magnitude should separate initial data for which they will remain stable from initial data in which they move to an energetically preferred state, such as a BS on another stable branch or pseudo-oscillaton.
The phenomenology of near-critical evolutions would then be an interesting object of study, with Type I critical solution appearing likely \cite{Choptuik:1996yg, Brady:1997fj}.
We plan to address this issue in future work.

\section*{Acknowledgments}
I am supported by the Cambridge Trust at the University of Cambridge. I am grateful for useful discussions with Seppe Staelens, Daniela Cors, Tamara Evstafyeva, and especially to my supervisor Ulrich Sperhake for his very helpful guidance in carrying out this research. 
Computations were done on
the CSD3, Fawcett and Swirles (Cambridge) and Cosma (Durham) clusters.

\appendix{}
\section{Convergence and Constraint Violations} \label{sec:convergence}
\begin{figure}[h!]
    \includegraphics[width=\linewidth]{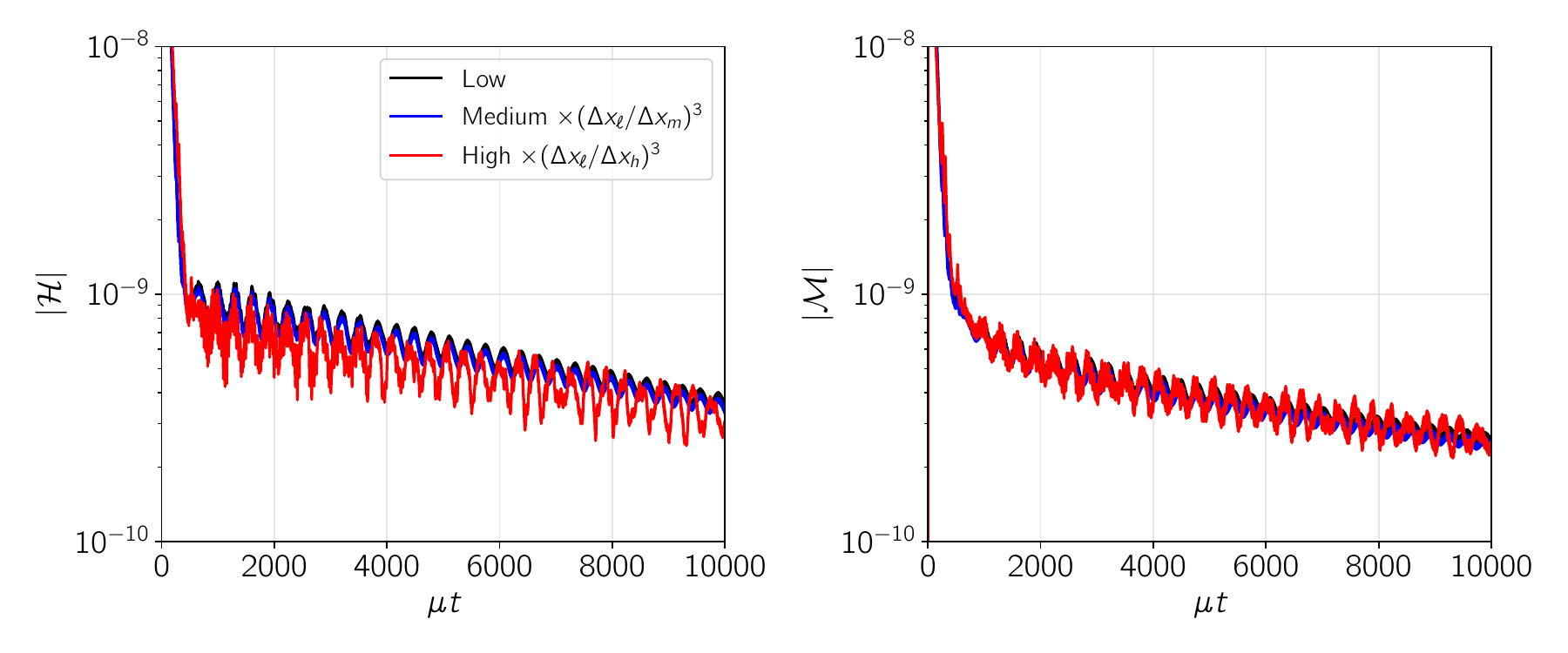}
    \caption{$L_2$ norms of the Hamiltonian (left) and momentum (right) constraints over time, for the 3+1 configuration \texttt{IIA} evolved at three resolutions $\Delta x_\ell = 1 / (12\mu),$   $\Delta x_m = 1 / (16\mu),$ and  $\Delta x_h = 1 / (20\mu)$ at the finest AMR level.
    We multiply the constraint violations at the two higher resolutions by the factors for third-order convergence, showing that our results are consistent with convergence between second and fourth order.
    \label{fig:aspherical_convergence}} 
\end{figure}

In this appendix, we present resolution studies demonstrating the numerical convergence of our dynamical evolutions.
Both {\sc grchombo} and {\sc sbse} track $L_2$ norms of the Hamiltonian and momentum constraint violations, given by
\begin{align}
    \mathcal{H} &= \mathcal{R} + K^2 - K^{mn}K_{mn} - 16\pi \rho , \\
    \mathcal{M}_i &= D_i K - D_m K^m_{\;\; i} + 8\pi j_i.
\end{align}
In Fig.~\ref{fig:aspherical_convergence}, we show the late-time convergence of the constraints towards zero for one of our aspherically perturbed evolutions, performed with {\sc grchombo}.
Note that convergence between second and fourth order is expected, as {\sc grchombo} uses fourth-order stencils and time integration, but incorporates second-order elements in the interpolation between AMR levels. 
We also remark that the decay in the extracted Noether charge converges to zero at approximately the same third order as the momentum constraint, suggesting that gravitational cooling is negligible for perturbations of this size.

\begin{figure}[h!]
    \includegraphics[width=\linewidth]{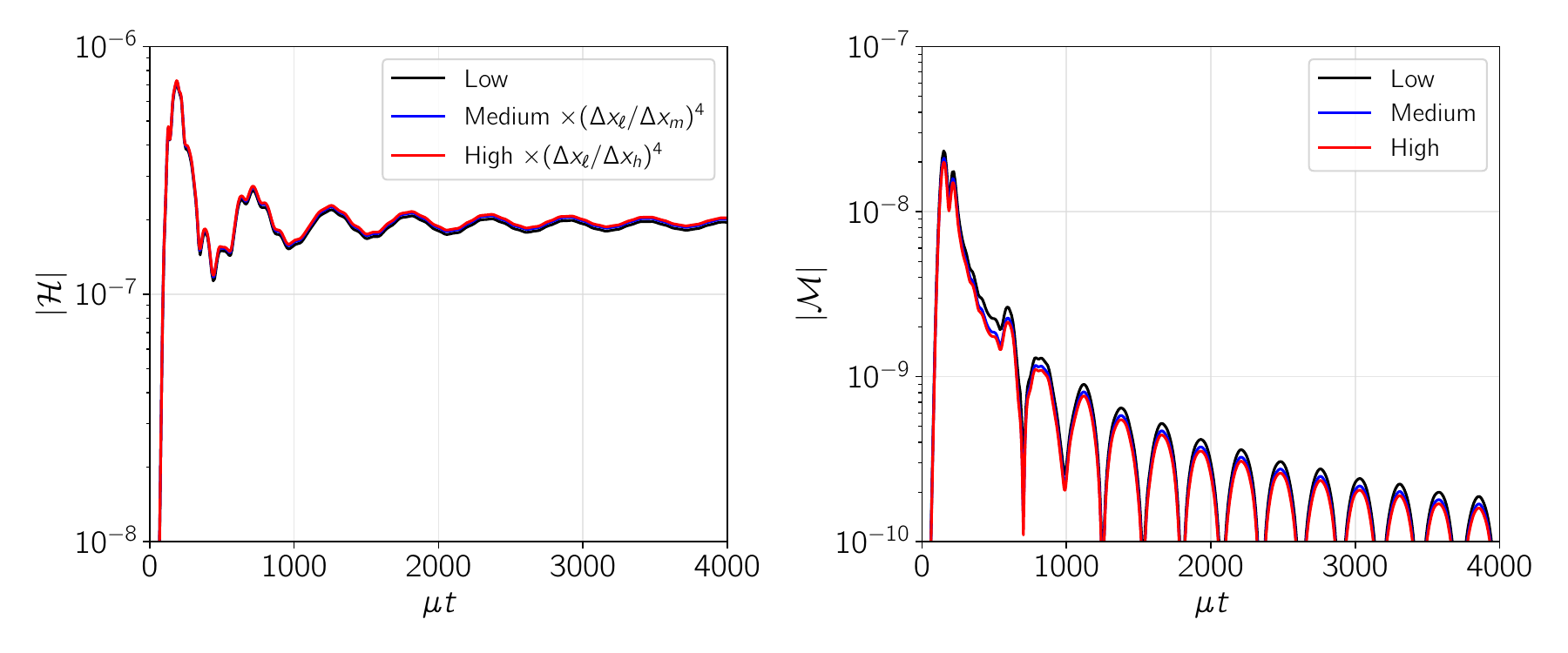}
    \caption{$L_2$ norms of the Hamiltonian (left) and momentum (right) constraints over time, for the pseudo-oscillaton-forming configuration shown in Fig.~\ref{fig:formation_fourier}, evolved at three resolutions $\Delta x_\ell = 1 / (16\mu),$   $\Delta x_m = 1 / (24\mu),$ and  $\Delta x_h = 1 / (32\mu)$.
    We multiply the Hamiltonian constraint violations at the two higher resolutions by the factors for third-order convergence, showing that our results are consistent with convergence at fourth order.
    The momentum constraint, on the other hand, is dominated by constraint violations propagating in from the outer boundary, so converges to a finite residual violation that decays in time.
    \label{fig:spherical_convergence}} 
\end{figure}

In Fig.~\ref{fig:spherical_convergence}, we repeat this for one of our {\sc sbse} evolutions resulting in the formation of a pseudo-oscillaton.
This time, the dominant Hamiltonian constraint violation cleanly decays to zero at the expected fourth order.
However, the momentum constraint violation is dominated by an error propagating inwards from the outer boundary, so converges not to zero but to a finite residual violation. 
Improving this would likely require the incorporation of a constraint-preserving boundary condition to {\sc sbse}, which we leave to future work.

In all, the convergence and small overall size of the constraint violations in both spherical and 3+1 evolutions provides additional support for the robustness of our results.

\bibliographystyle{iopart-num}
\bibliography{ref}

\end{document}